\documentclass[sn-apa]{sn-jnl}% APA Reference Style
\jyear{2022}%
\usepackage{amsfonts}       % blackboard math symbols
\usepackage{nicefrac}       % compact symbols for 1/2, etc.
\usepackage{microtype}      % microtypography
\usepackage{xcolor}         % colors

\usepackage{amsmath}
\usepackage{graphicx}
\usepackage{tabu}
\usepackage{multirow}
\usepackage{bm}
\usepackage{times}

\begin{document}
	
\title[Inequity aversion reduces travel time in the traffic light control problem]{Inequity aversion reduces travel time in the traffic light control problem}

\author[1]{\fnm{Mersad} \sur{Hassanjani}}\email{mersadhassanjani@gmail.com}
\equalcont{These authors contributed equally to this work.}

\author*[1]{\fnm{Farinaz} \sur{Alamiyan-Harandi}}\email{farinaz.alamiyan@gmail.com}
\equalcont{These authors contributed equally to this work.}

\author[2]{\fnm{Pouria} \sur{Ramazi}}\email{p.ramazi@gmail.com}

\affil*[1]{\orgdiv{Department of Electrical \& Computer Engineering}, \orgname{Isfahan University of Technology}, \orgaddress{\city{Isfahan}, \postcode{84156-83111}, \country{Iran}}}

\affil[2]{\orgdiv{Department of Mathematics \& Statistics}, \orgname{Brock University}, \orgaddress{\street{St. Catharines, ON L2S 3A1}, \country{Canada}}}

\abstract{
The traffic light control problem is to improve the traffic flow by coordinating between the traffic lights. Recently, a successful deep reinforcement learning model, CoLight, was developed to capture the influences of neighboring intersections by a graph attention network. We propose IACoLight that boosts up to 11.4\% the performance of CoLight by incorporating the Inequity Aversion (IA) model that reshapes each agent's reward by adding or subtracting advantageous or disadvantageous reward inequities compared to other agents. Unlike in the other applications of IA, where both advantageous and disadvantageous inequities are punished by considering negative coefficients, we allowed them to be also rewarded and explored a range of both positive and negative coefficients. Our experiments demonstrated that making CoLight agents averse to inequities improved the vehicles' average travel time and rewarding rather than punishing advantageous inequities enhanced the results.
}


%%\pacs[JEL Classification]{D8, H51}

%%\pacs[MSC Classification]{35A01, 65L10, 65L12, 65L20, 65L70}

\maketitle

\section{Introduction}
\label{S.Introduction}

The problem of traffic light control is to coordinate between intersections by controlling their traffic lights to improve traffic flow. This problem remains as one of the greatest challenges in the $21$st century \citep{qadri2020state}. To tackle this challenge, researchers have taken various approaches such as the coordinated method modifying the start time of the green lights between the consecutive intersections \citep{koonce2008traffic}, the optimization technique minimizing the vehicles' travel time under certain traffic flow assumptions \citep{diakaki2002multivariable}, and the models applying perimeter control to handle transferring flows between regions of a city \citep{kouvelas2017linear, kouvelas2015feedback}. In addition to conventional approaches, the problem was recently tackled with Reinforcement Learning (RL) methods \citep{qadri2020state}. RL is a promising machine-learning framework where an agent interacts within a given environment by applying \textit{action}s and receiving signals, which are interpreted as \textit{reward}s and \textit{punishment}s. Via the interactions, the agents learn an optimal \textit{policy}, a probability distribution over the available actions that maximizes the total obtained rewards for each visited environment \textit{state} \citep{sutton1998introduction,alamiyan2018new,rasheed2020deep}.	 

Encompassing several intersections, the traffic light control problem requires several actions to be executed at the same time. Hence, often the Multi-Agent (MA) extension of RL, i.e., MARL,  is used for this problem.
In the MARL setting, several agents coexist in an environment with multiple intersections. Each agent is responsible to control the traffic flow of one intersection by scheduling the traffic lights and learns a policy from its own observations to minimize the average queue length on all lanes over all intersections. This multi-agent setting resulted in a conflict of interest in using intersections as the common resources. 

The CoLight \citep{wei2019colight} is a state-of-the-art model to enable cooperation of traffic signals in the traffic light control problem. The authors utilized a graph attention network \citep{velivckovic2018graph} that represents the observations of the neighboring intersections as an overall summary and enables the agents to learn the neighboring influence on their under-control intersection. Each agent managed the traffic flow of an intersection by using a Deep Q Network (DQN) structure in the CityFlow simulator, that is an open-source traffic simulator designed for large-scale traffic scenarios \citep{zhang2019cityflow}.	

Recently, the Inequity Aversion (IA) model \citep{hughes2018inequity} was introduced to improve the performance of MARL algorithms by manipulating the agents' rewards based on the envy and guilt notions. Each agent compares its reward with that of each of its fellows. An \textit{advantageous inequity} happens when the agent earns more in a pairwise comparison, and a \textit{disadvantageous inequity} happens otherwise. Both cases are punished by subtracting the difference from the agent's own reward, capturing the guilt and envy notions, respectively.

Nevertheless, advantageous and disadvantageous inequities may be considered as a reward, rather than a punishment, by adding a positive scale of differences to the agent's own reward. Yet this has not been thoroughly investigated in the literature. To the best of our knowledge, all previous research on applying the IA model to MARL has used inequities as punishments. It hence remains an open question whether and how much rewarding the inequities can improve the performance of the IA model. 
This is partly due to the great computational costs associated with obtaining the results for a single pair of coefficients for the inequities. In the original work, a limited search over the parameter space is performed to find the best range.   

Our goal is \emph{(i)} to investigate the effectiveness of reshaping rewards in MARL in the traffic light control problem by using the IA model and \emph{(ii)} to investigate the effect of different coefficients of the advantageous and disadvantageous inequities. To this end, we incorporate the IA model into the training process of the CoLight model \citep{wei2019colight} and introduce the IACoLight model. The obtained results are compared with the CoLight's result under the same simulated environments. We also investigate the positive and negative ranges of the IA model's hyperparameters to analyze the potential benefits achieved from their different combinations in the traffic light control problem. 

The remainder of this paper is organized as follows: Section \ref{S.RL_Traffic} reviews some RL methods that have been recently introduced to control traffic lights. Section \ref{S.IACoLight} defines the MARL problem and the proposed IACoLight model.
The experimental results are presented in Section \ref{S.Experiments} and are discussed in Section \ref{S.Discussion}.
\section{RL methods to control traffic lights}
\label{S.RL_Traffic}

To apply RL methods in traffic signal control, researchers considered various traffic scenarios and scales from controlling the traffic flow of a single intersection to several ones in crowdy cities. 

To control the traffic lights of a single intersection in the Simulation of Urban MObility (SUMO) simulator \citep{SUMO2018}, the DQN structure of the IntelliLight method \citep{wei2018intellilight} was trained using real data collected from surveillance cameras. The learned policy was evaluated not only based on the quantitative evaluation metrics including the weighted sum of the average of rewards, queue lengths, durations, and delays but also by a qualitative assessment
of the pattern learned for light switching, because some policies might have the same reward but one may be more suitable for real-world practice.

According to an analogy that ``when humans attempt to master a skill, they often refer to expert knowledge'', the authors of the DemoLight \citep{xiong2019learning} utilized the Self-Organizing Traffic Light Control \citep{cools2013self} to make the agents learn from an expert. They applied the Advantage Actor-Critic (A2C) algorithm \citep{mnih2016asynchronous} to speed up the learning process in the CityFlow simulator. The results of the DemoLight were evaluated by the travel time metric and demonstrated a more efficient exploration of this approach in comparison with other previous methods.  

The PressLight algorithm was introduced in \citep{wei2019presslight} where the reward function was defined according to the Max Pressure method \citep{lioris2016adaptive} that minimizes the overall travel time for the whole network by minimizing the pressure of each intersection in the network. A DQN was trained to control the traffic lights of several intersections in the CityFlow simulator using both synthetic and real-world traffic data. The performance excelled under heavy traffic. 

MetaLight \citep{zang2020metalight} was another approach where the gradient-based meta-learning algorithm \citep{finn2017model} was applied in the CityFlow simulator to speed up the learning process and extend generalization so that the knowledge gained in previous traffic scenarios could be used in new ones. This method also improved the FRAP\footnote{A model that is invariant to symmetric operations like Flipping and Rotation and considers All Phase configurations.} model \citep{zheng2019learning}, that was based on the DQN structure to train the agents. Individual and global level adaptions were used to apply meta-learning on the off-policy RL method of the MetaLight. MetaLight was tested on four real-world datasets.    

MPLight method \citep{chen2020toward} was another DQN-based method that employed a decentralized DQN structure which used the concept of pressure in traffic and utilized parameter sharing to control all intersections. Although the reward function and state definition were the same as the ones introduced in the PressLight method, FRAP was used in MPLight instead of a simple DQN as the base model. Scalability, coordination, and data feasibility were the three main problems tested in the experiments with more than $1,000$ traffic lights using the CityFlow simulator.  
\section{The IACoLight setup}
\label{S.IACoLight}

We define the traffic light control problem in the following MARL setting. Consider an \emph{environment} in the form of an urban region consisting of several roads with $N$ intersections, each having a number of traffic lights controlled by a single agent, resulting in a total of $N$ agents. At each time step $t$, each agent $k$ makes the partial observation $o^k_t$ that consists of \emph{(i)} the number of vehicles waiting on each of the lanes of the intersection and \emph{(ii)} the light of which lanes of the intersection are green. The observation $o^k_t$ is represented by some of its features extracted by a layer of Multi-Layer Perceptron (MLP). The representation is referred to as the environment's \textit{local state} $s^k_t$ and is communicated to the agent's neighboring intersections via graph attentional networks \citep{velivckovic2018graph}. These networks prepare a comprehensive summary of the intersection's
neighborhood indicating the importance of the received information from each neighbor. 

\begin{center}
	\begin{figure}[!h]
		\centering	
			\includegraphics[width=0.5\textwidth]{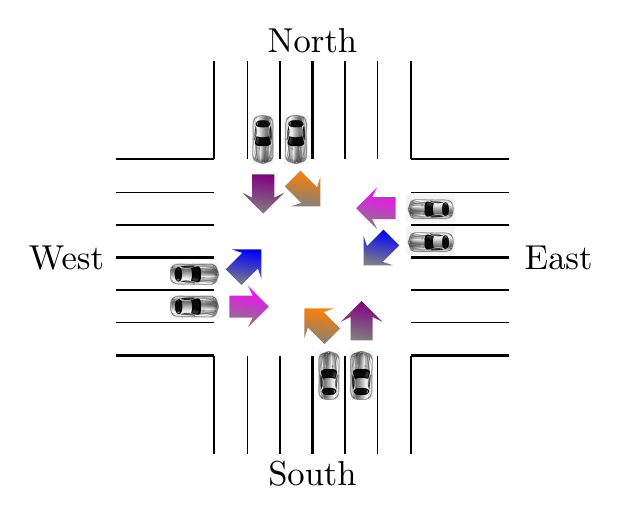}\caption{\textbf{An intersection with four phases of light control.} Pairs of same-color arrows show a phase of this intersection.} 
		\label{fig.intersection}
	\end{figure}
\end{center}

Each agent executes an \textit{action} that determines the intersection's \textit{phase} specifying the light of which two traffic movements of the intersection will be green during the next time interval $\Delta t$. For example, consider an agent that controls a typical intersection with four entering approaches marked with the main directions east, north, west, and south (Figure \ref{fig.intersection}). The vehicles in each entering approach should choose from one of the three lanes: the right lane to turn right, the left lane to turn left and the middle lane to go straight ahead. For this intersection, a standard phase turns the light green for vehicles in the middle lanes of two opposite entrances. 

The available actions for agent $k$ form the
\textit{action set} $\mathcal{A}^k=\{a_1,\ldots,a_{m}\}$.
Agent $k$ selects an action $a^k_t$ by using a \textit{policy} $\pi^k$, that is a probability distribution over the agent's action set. 
By executing the joint action $\bm{a}_t=[a^1_t,...,a^k_t,...,a^N_t]$ at global state $s_t \in \mathcal{S}$, the features extracted  from all agents' observations, the environment transfers to the global state $s_{t+1}$ according to a transition distribution $\mathcal{T}(s_{t+1} \lvert s_{t},\bm{a}_{t})$, resulting in the reward $r^k_{t+1}$ to each agent $k$. Aiming to minimize the vehicles' travel time, we define the reward of each agent $k$ based on the average length of the queue formed in the incoming lanes of every direction of the agent's intersection as $r^k_{t}=-\frac{1}{d}\sum_{l}u_{t}^{k,l}$ 
where $u_{t}^{k,l}$ is the queue length of lane $l$ of intersection $k$ at time $t$ and $d$ is the number of the intersection's entrances.

Each agent $k$ uses its rewards to compute a \textit{state-action value function} $Q^{\pi^k}(s^k_t,a^k_t)$ for all local states $s^k_t$ and actions $a^k_t\in \mathcal{A}^k$, where $Q^{\pi^{k}}(s^{k}_{t},a^k_t)$ approximates the \textit{return} $R^{k}_{t}=\sum_{i=0}^{\infty}\gamma^{i}r^{k}_{t+i+1}$, $0\leq\gamma\leq1$, that is an estimation of the cumulative \textit{$\gamma$- discounted rewards} over all local states visited in the future by applying action $a^k_t$ on local state $s^k_t$ and following policy $\pi^k$. Agent $k$ selects the action with maximum $Q^{\pi^k}(s^k_t,a^k_t)$ at each time step $t$. 

\begin{center}
	\begin{figure}[!h]
		\centering
		\resizebox{\columnwidth}{!}{
		\includegraphics{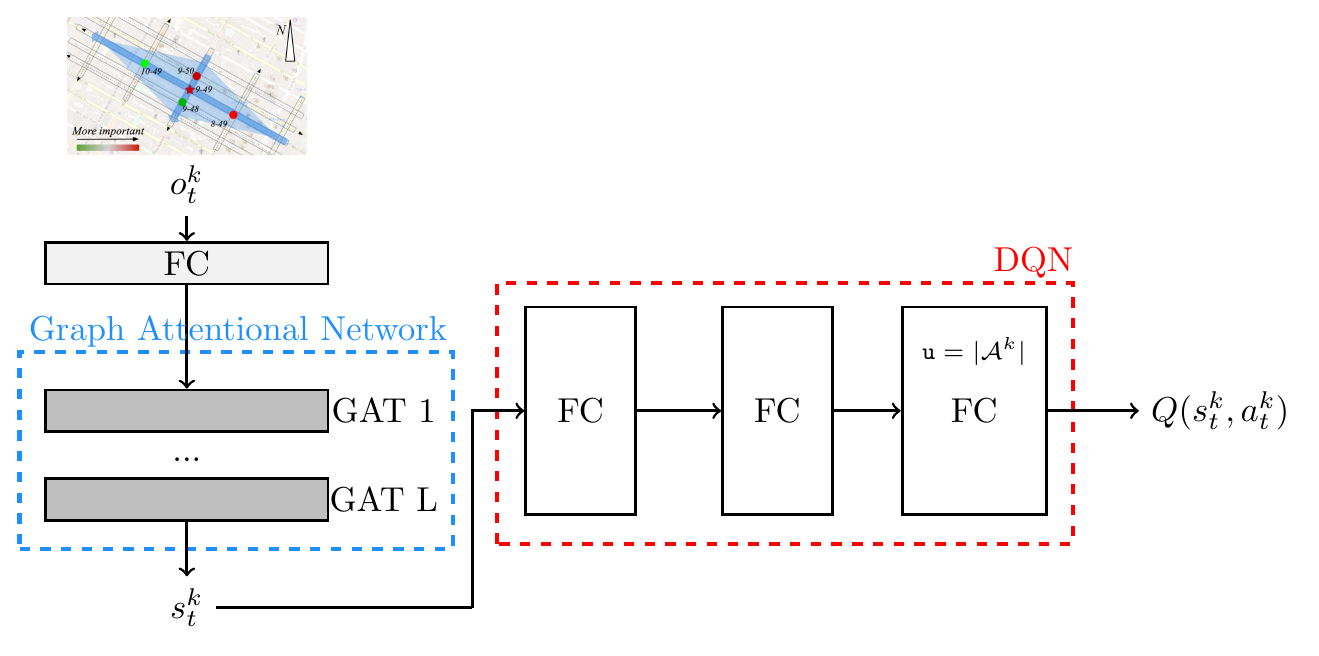}}
		\caption{\textbf{The agents' network structure.} The inputs of each agent $k$'s DQN is the local state $s^k_t$ that is composed of the representing features provided by the output of the graph attentional network that is utilized for a graph-structured environment where a summary of each graph node's neighborhood is prepared by embedding information from the neighbors and their importance to the central node. The $L$ graph attention layers of this network are illustrated by GAT. The DQN structure learns the state-action value function. Fully connected layers are indicated by FC, and the parameter \texttt{u} determines the number of their neurons. The graph attentional network and DQN structures are taken from \citep{wei2019colight}.} 
	    \label{fig.NetStructure}
	\end{figure}
\end{center}

Here similar to the CoLight model \citep{wei2019colight}, a DQN (Figure \ref{fig.NetStructure}) \citep{mnih2015human} is used as $Q^{\pi^k}(s^k_t,a^k_t)$; it takes the visited local state $s^k_t$ as the input and estimates the state-action value of applying each available action of agent $k$ in the local state $s^k_t$ as the output. The parameters of the DQN, denoted by $\bm{\theta}$, are learned iteratively by minimizing the following loss function:
\begin{equation}\label{equ.Q_update_formula}
\begin{aligned}
\mathcal{L} (\bm{\theta}_n) = E \left[ {\left(r^k_t + \gamma {\max_{a^k_{t+1}}} Q^{\pi^k}({s^k_{t+1}},{{a}^k_{t+1}}; \bm{\theta}_{n-1}) - Q^{\pi^k}(s^k_t,a^k_t; \bm{\theta}_{n})\right)}^2 \right] ,
\end{aligned}
\end{equation}
where $\bm{\theta}_n$ is the parameter vector of the DQN's neural networks in the $n$th iteration of the learning process. The loss is measured as the differences between the predicted and actual (target) $Q$ values. Term $Q^{\pi^k}(s^k_t,a^k_t; \bm{\theta}_{n})$ is the predicted $Q$ value and term $r^k_t + \gamma {\max_{a^k_{t+1}}} Q^{\pi^k}({s^k_{t+1}},{{a}^k_{t+1}}; \bm{\theta}_{n-1})$ is the target $Q$ value that computed by using the result of DQN in presence of $\bm{\theta}_{n-1}$, the DQN's parameters in the previous iteration of the learning process.
To implement the IACoLight model, we used the linear reward function $r^k_t = \alpha e^k_t + \beta i^k_t $ where $\alpha$ and $\beta$ are constant scalars, $e^k_t$ is the extrinsic reward that agent $k$ receives from the environment, and $i^k_t$ is the intrinsic reward that agent $k$ computes according to the IA model \citep{hughes2018inequity}:
\begin{equation}\label{equ.Inequity_Aversion_reward}
\begin{aligned}
i_{t}^{k} = -\frac{\alpha_{k}}{N-1}\sum_{j\neq k} \max(w^{j}_{t} - w^{k}_{t}, 0) - \frac{\beta_{k}}{N-1}\sum_{j\neq k} \max(w^{k}_{t} - w^{j}_{t}, 0),
\end{aligned}
\end{equation}
where $\alpha_{k}$,$\beta_{k}\in\mathbb{R}$ are adjustable parameters and $w^j_t$ is a temporary memory of the extrinsic reward occurrence \citep{hughes2018inequity}: 
\begin{equation}\label{equ.Inequity_Aversion_e}
\begin{aligned}
w^{j}_{t} = \gamma \lambda w^{j}_{t-1} + e_{t}^{j}  \;\;\;\;\;\;\;\forall t\geq 1, \; w^j_0 = 0,
\end{aligned}
\end{equation}
and where $\lambda\in[0,1]$ is a trace-decay hyper-parameter.
In equation (\ref{equ.Inequity_Aversion_reward}), the terms $\max(w^{k}_{t} - w^{j}_{t}, 0)$ and $\max(w^{j}_{t} - w^{k}_{t}, 0)$ are the advantagous and disdvantagous inequities of agent $k$ against one of its fellows, agent $j$, respectively. 

\section{Experiments}
\label{S.Experiments}

\subsection{Experiment setup}
\label{S.Experiments.Experiment_Setup}

We used the same experiment setup as that of CoLight \citep{wei2019colight} where several environments were generated based on the CityFlow simulator. The authors collected data by analyzing the trajectories of vehicles in captured images of roadside cameras in the Chinese cities, Hangzhou and Jinan, as well as New York city in the United States. In these environments, several vehicles move from various origins to their destinations while encountering traffic lights. The green signal of each traffic light is accompanied by a yellow three-second signal and a minimum of two seconds red signal. The authors extracted traffic information such as the number of vehicles passing through the intersections during a single day. They also used some artificially generated data.

In the Hangzhou dataset, the data is collected from roadside surveillance cameras existing in 16 intersections in Gudang Sub-district Hangzhou, China. The mean and standard deviation of the arrival rate (vehicles/$300s$) in this dataset are $526.63$ and $86.70$, respectively.
In the Jinan dataset, the extracted data is associated with roadside cameras located on $12$ intersections in Dongfeng Sub-district, Jinan, China. The mean and standard deviation of the arrival rate (vehicles/$300s$) in this dataset are $250.70$ and $38.21$, respectively.

We conducted our experiments on the environment that has been created by using the Hangzhou and Jinan datasets\footnote{https://traffic-signal-control.github.io}, which consists of one $4\times4$ and another ($3\times4$) grid with $16$ (resp. $12$) intersections. Each agent controls an intersection and sends its information to its four neighbors located at its four main sides. To learn appropriate policies in these environments, each experiment included $100$ episodes, each containing $1440$ samples, which is the maximum amount of data that can be created in an intersection during a day. 

We compared IACoLight with CoLight \citep{wei2019colight} as a baseline. For the CoLight method, we set the number of heads in the attention mechanism to $5$. 
For IACoLight, we used the advantageous type of IA model where $\alpha$ is zero and $\beta$ is set to $0.05$. 
To evaluate the performance of each method, the time distance between entering and leaving the grid for each vehicle was measured and their average was used as the vehicle travel time metric. 
We executed each method $5$ times, with different traffic flow sampled from the main datasets, and obtained the average of the evaluation metric.	

Since the IA model is sensitive to hyperparameters, we ran a sweep over $\alpha$ and $\beta$ parameters in order to calibrate them correctly in the traffic light control problem and also analyze their synergic effect.
Here, to investigate the effect of agents' aversion to each of the advantageous and disadvantageous inequities, the range of $-1$ to $1$ with increments of $0.2$ for both $\alpha$ and $\beta$ was tested, resulting in a total of $11\times11 = 121$ cases, each repeated $3$ times and for a length of $100$ episodes. This resulted in a total of $363$ experiments.
Each experiment took around $5$ hours and was performed on a Linux server with $16$ CPUs and $120$G RAM.  
				
\subsection{Experimental results}
\label{S.Experiments.Experimental_Results}

\begin{center}
	\begin{figure}[!h]
		\centering			
		\resizebox{\columnwidth}{!}{
			\extrarowsep=_3pt^3pt			
			\begin{tabu}to\linewidth{c}
			{\includegraphics[width=\columnwidth ]{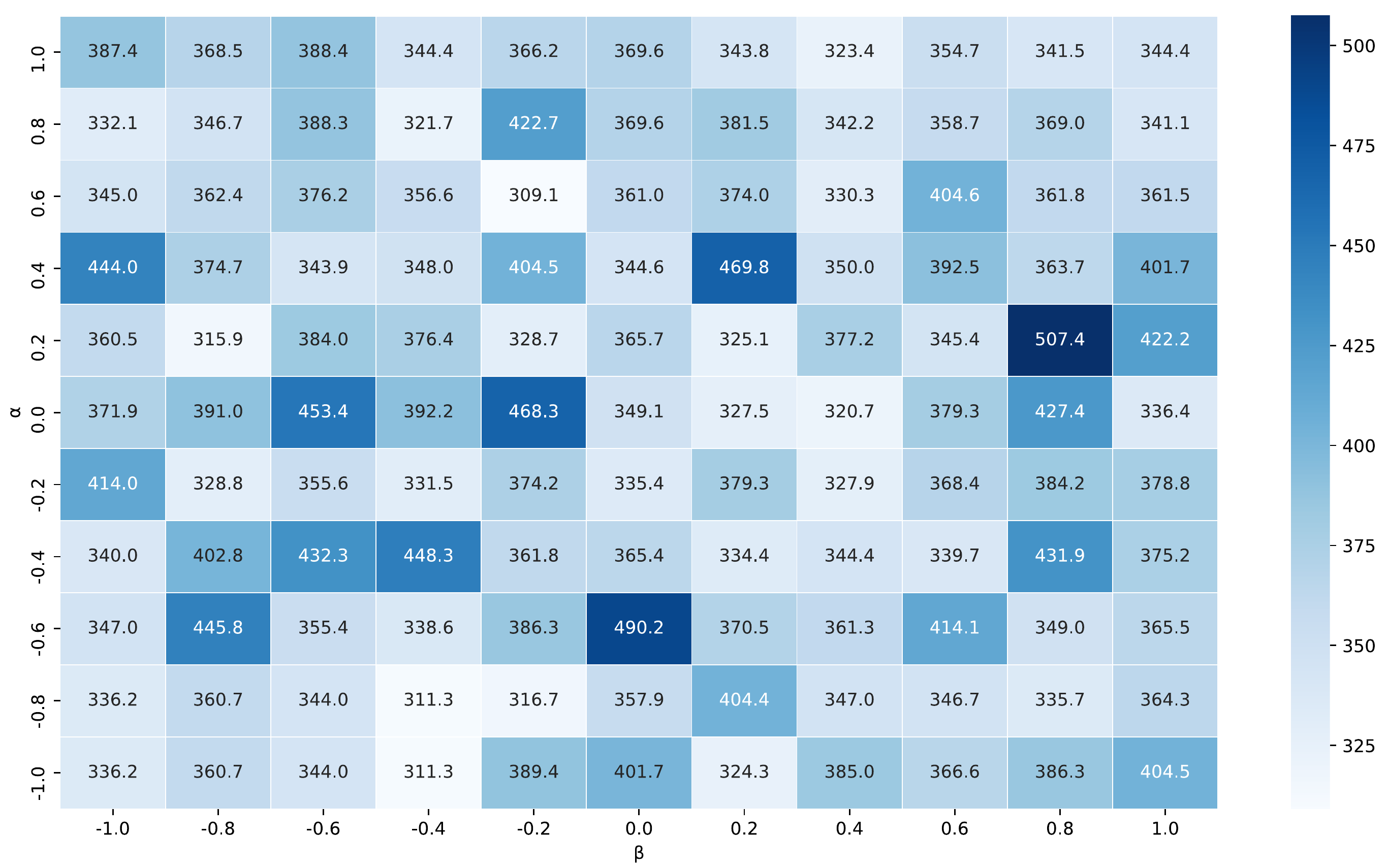}} 			
		\end{tabu}}
		\caption{\textbf{The results of the average travel time over the last $\textbf{20}$ episodes achieved by the $\textbf{3}$ experiments done for each $\bm{\alpha}$ and $\bm{\beta}$ combination of the IACoLight model using the Hangzhou dataset.} The result of CoLight is obtained by $\alpha=0, \beta=0$.}
		\label{fig.AlphaBetaSearch}
	\end{figure}
\end{center}

According to the result of tuning the $\alpha$ and $\beta$ hyperparameters of IACoLight in the Hangzhou dataset, using both inequities of the IA model with tuned coefficients in the IACoLight model outperforms all other methods (Table \ref{Table.expriment_results}). The best value of the average travel time computed for the last $20$ episodes as a measure of the final learned
performance was $309.1$ that belonged to $\alpha=0.6$ and $\beta=-0.2$. It was $11.4\%$ better than the state-of-the-art CoLight model. This performance is achieved when disadvantageous inequities were punished ($\alpha >0$) but advantageous inequities were rewarded ($\beta <0$), which is more than the common setup in the literature where only disadvantageous inequities are used and punished ($\alpha >0, \beta =0, 1.3\%$), or only advantageous inequities are used and punished ($\alpha=0, \beta >0, 8.1\%$). Note that these differences are significant in the context of traffic flow control. The next best $3$ performance values were $311.3$, $315.9$, and $316.7$. All of these values are lower than $325.1$, the performance of the common setup in the literature where both advantageous and disadvantageous inequities were punished ($\alpha,\beta>0$).	 
These results demonstrated that the lower mean travel time for vehicles was obtained when $\beta$ was negative (Figure \ref{fig.AlphaBetaSearch}). 

\begin{center}
	\begin{figure}[!h]
		\centering			
		\resizebox{\columnwidth}{!}{
			\extrarowsep=_3pt^3pt			
			\begin{tabu}to\linewidth{c}
				{\includegraphics[width=\columnwidth]{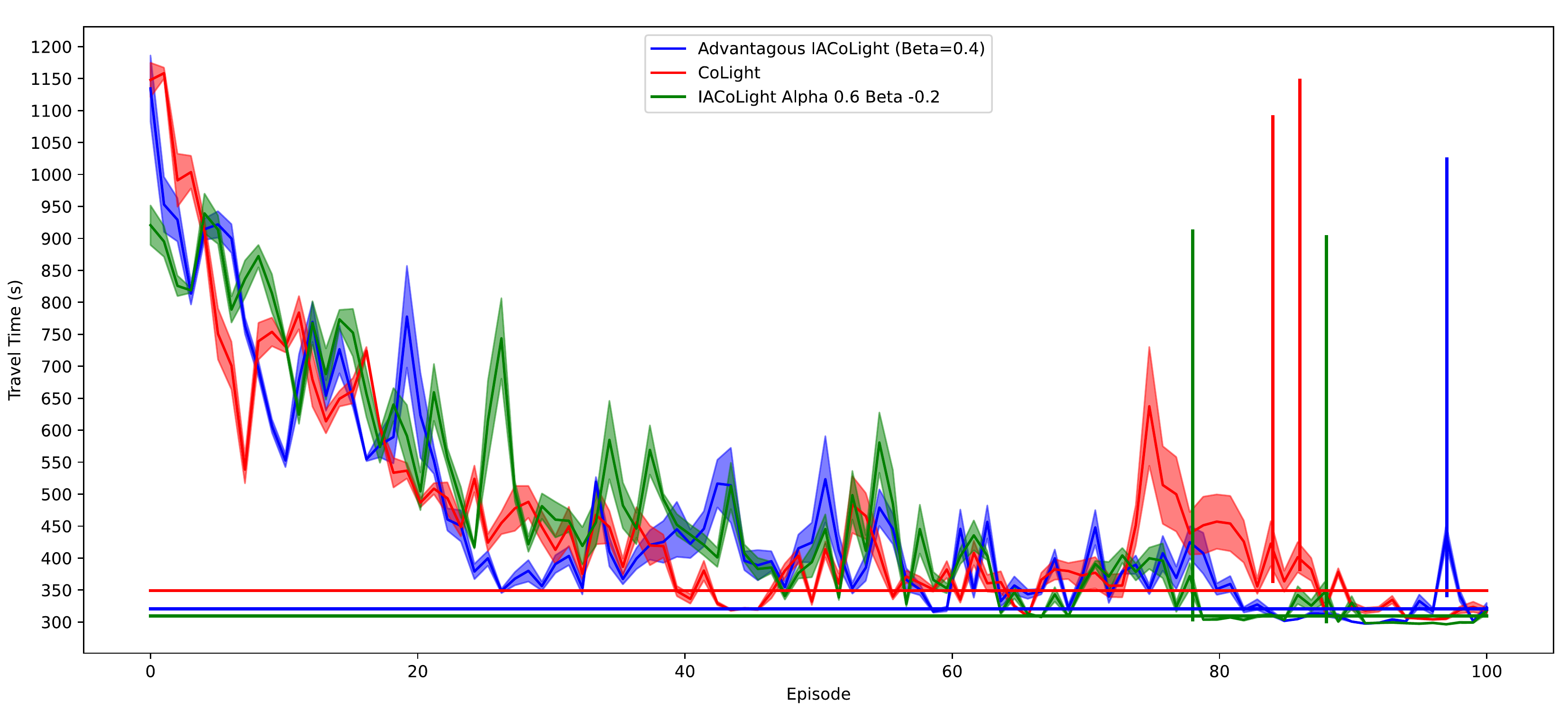}}  
		\end{tabu}}
		\caption{\textbf{Comparison of the mean travel times of the methods on the Hangzhou dataset.} Each point of these curves represents the average delay time of all $16$ agents over $3$ individual experiments. The shadows are the bands of the confidence interval obtained by estimating the unbiased variance of the travel times. The vertical lines illustrated with the same color as each method's curve represent the position where the method converges. The closer this line is to the vertical axis, the faster the method's convergence. To draw these lines, we calculated the average travel time obtained in the last $20$ episodes of each method as a threshold. Then we checked if there is an episode in the range of $50$ and $100$ that the maximum travel time computed for each episode from that episode to the end episode is less than $1.2$ and $1.1$ times the threshold. If the episodes with the first and second conditions were found, we determined the first and second vertical lines at those episode numbers, respectively. The horizontal lines indicate the average of the last $20$ episodes of each run, where a lower line means less travel time. The exact values of this metric are $349.1$, $320.7$ and $309.1$ for CoLight, the advantagous IACoLight and IACoLight with best $\alpha$ and $\beta$ combination, respectively.
		}
		\label{fig.Reward_Comparison_Hangzhou}
	\end{figure}
\end{center}

\begin{table}[h]
	\begin{center}
		\begin{minipage}{\textwidth}
			\caption{\textbf{Statistical results for the Hangzhou dataset.}}\label{Table.expriment_results}
			\begin{tabular*}{\textwidth}{@{\extracolsep{\fill}}lccc@{\extracolsep{\fill}}}
				\toprule				
				 Methods & Episode & Performance & Convergence  \\
				 & (From $0$ to $100$) & (Second) & (Episode numbers)  \\
				\toprule
				CoLight  & $466.7$ & $349.1$ & $84,86$  \\
				\midrule	
				Disadvantageous IACoLight ($\beta = 0$) &   &   &   \\	
				$\alpha = 0.4$  & $474.5$ & $344.6$  & $82,95$   \\	
				$\alpha = -0.2$ & $461.2$ & $335.4$  & $87,95$   \\
				\midrule
				Advantageous IACoLight ($\alpha = 0$)  &   &   &   \\		
				$\beta = 0.4$  & $\textbf{458.4}$ & $320.7$  & $97,97$  \\
				$\beta = -1$  & $481.3$ & $371.9$ & $73,96$  \\
				\midrule
				IACoLight ($\alpha,\beta>0$)&   &   &   \\
				$\alpha = 0.2 , \beta = 0.2$ & $460.1$ & $325.1$  & $74,99$   \\
				\midrule
				IACoLight (Exhaustive search)&   &   &   \\
			    $\alpha = 0.6 , \beta = -0.2$ & $473$ & $\textbf{309.1}$  & $\textbf{78},\textbf{88}$   \\
				\botrule			
			\end{tabular*}
			\footnotetext{Note: The first numerical column displays the average travel time obtained by whole vehicles in episodes from $0$ to $100$ over $3$ individual experiments of CoLight and IACoLight for the Hangzhou dataset. The second and third numerical columns indicate the final learned performance and the convergence thresholds, respectively. The $\alpha$ and $\beta$ values are reported for the best results of different types of IACoLight considering positive and negative ranges of these coefficients.
		 }					
		\end{minipage}
	\end{center}
\end{table}
In the Hangzhou dataset, the convergence speed of IACoLight with best $\alpha$ and $\beta$ combination  was slightly higher, but CoLight and the advantageous IACoLight, with the commonly used $\alpha$ and $\beta$ values in the literature, had a lower distance between the two convergence indexes (the vertical lines in Figure \ref{fig.Reward_Comparison_Hangzhou}). The final learned performance demonstrated that IACoLight with best $\alpha$ and $\beta$ combination outperformed the CoLight and the advantageous IACoLight methods (the horizontal lines in Figure \ref{fig.Reward_Comparison_Hangzhou}). It had a $11.4\%$ and $3.6\%$ lower travel time compared to CoLight and the advantageous IACoLight over the last of $20$ episodes.

\begin{center}
	\begin{figure}[!h]
		\centering			
		\resizebox{\columnwidth}{!}{
			\extrarowsep=_3pt^3pt			
			\begin{tabu}to\linewidth{c}
				{\includegraphics[width=\columnwidth ]{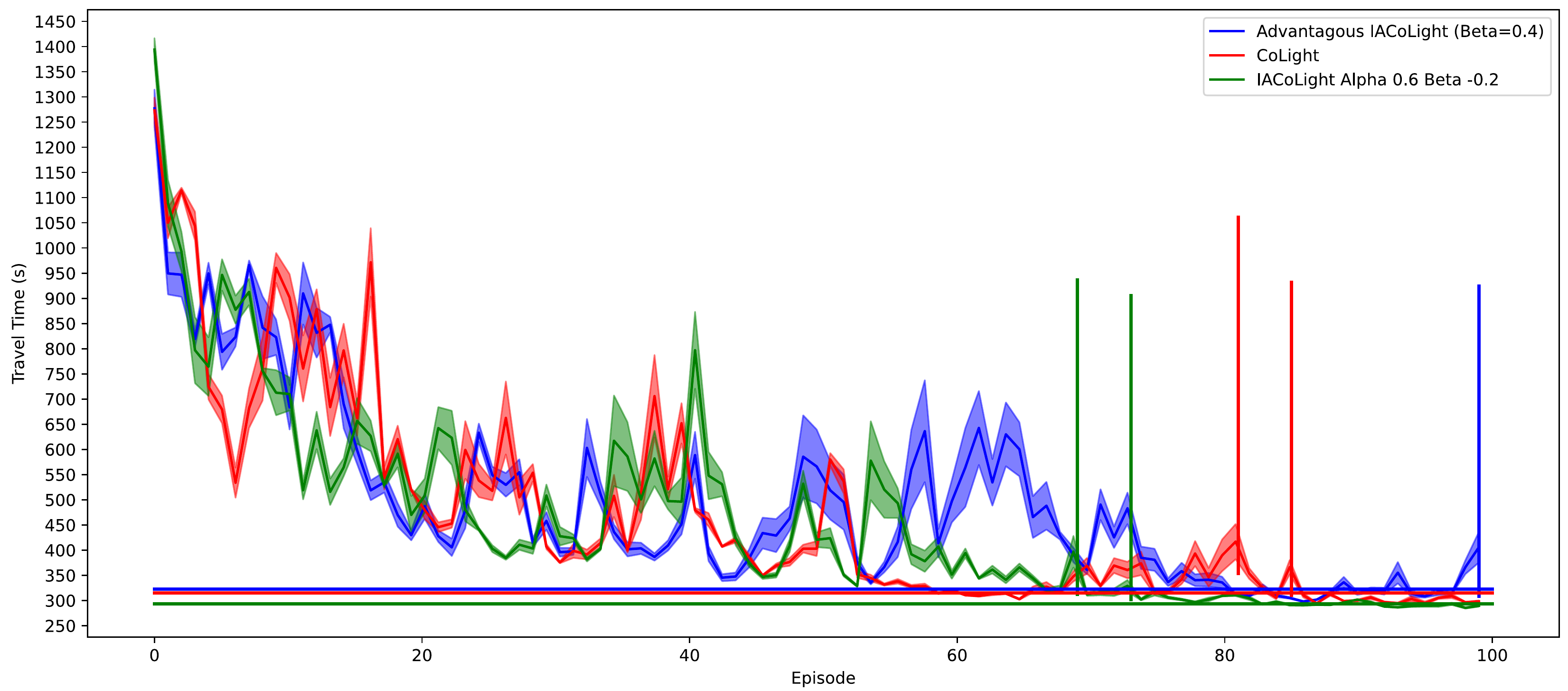}}  
		\end{tabu}}
		\caption{\textbf{Comparison of the mean travel times of the methods on the Jinan dataset.} The same setup as that of Figure \ref{fig.Reward_Comparison_Hangzhou} is used. The exact values of the horizontal lines are $314.9$, $322.7$ and $292.9$ for CoLight, the advantagous IACoLight and IACoLight with best $\alpha$ and $\beta$ combination, respectively. The convergence threshold pairs of these methods are $(81,85)$, $(99,99)$ and $(69,73)$, respectively.}
		\label{fig.Reward_Comparison_Jinan}
	\end{figure}
\end{center}

In the Jinan dataset, IACoLight with best $\alpha$ and $\beta$ combination reached both convergence threshold sooner. In addition, its final learned performance had $7\%$ and $9.2\%$ lower travel time compared to CoLight and the advantageous IACoLight (Figure \ref{fig.Reward_Comparison_Jinan}).
\section{Discussion}\label{S.Discussion}

We boosted the performance of agents trained based on the state-of-the-art CoLight method in the MARL setting of the traffic light control problem by utilizing the IA method inspired by the envy and guilt feelings. We conducted the experiments on real data from Hangzhou and Jinan cities and compared the result of IACoLight with CoLight. The comparison results showed that IACoLight reduces the average travel time of the vehicles.

The recommended intervals for the $\alpha$ and $\beta$ parameters of the IA model in the RL literature were as follows: $\alpha > \beta$ and $\beta \in [0,1]$ \citep{hughes2018inequity,yang2020learning,jiang2019learning}. These settings were adopted on the basis that people are averse to inequities against themselves more than inequities against others. So they punish themselves more when they face disadvantage inequities.
For the first time, we performed an exhaustive search over these parameters of the IA model. Unlike the common practice of using positive values for both $\alpha$ and $\beta$ in the literature, we found that negative values of $\beta$ can yield a higher performance. This corresponds to the case where advantageous inequities are rewarded. 

The superior performance of IACoLight may be explained by the positive and negative coefficients of the reward inequities. A positive value of the $\alpha$ parameter implies punishing the agent when ``feeling'' envy, causing the agent to act in order to improve its performance. However, negative values of $\alpha$ encourage the agent to continue its current behavior. On the other hand, assigning positive values to the $\beta$ parameter induces the feeling of guil whereas negative values encourage the agent by increasing its received reward, which may be interpreted as the feeling of pride. 

Now, the common practice is to take both $\alpha$ and $\beta$ positive, implying that both the agent who earned less and the one who earned more punish themselves and try to change their behaviour. This may lead to a population where all indivdiuals earn the same reward, but with the possible cost of loosing the valuable actions taken by the highest-earners. However, when $\beta$ is negative, only the less-earning agent punishes itself and tries to change its behaviour. The higher-earning continues its rewarding actions. Which of these two or other combinations of $\alpha$ and $\beta$ perform best? It seems to depend on the environment, and remains as a future work to be further investigated. 

The best advantageous type of IACoLight , i.e., $\alpha = 0$, had a positive $\beta$ value ($0.4$) and the best disadvantageous type of IACoLight, i.e., $\beta = 0$, had a negative $\alpha$ value ($-0.2$). Namely, if only one type of inequity is considered, it is better to feel guilty against the advantageous inequities and make the agent punish itself using larger positive values compared the case when both inequities are considered. On the other hand, it is better to reward the agent against the disadvantageous inequities using lower negative values compared the case when both inequities are considered.

It is noticeable that unlike the problems discussed in the IA model in the RL literature \citep{hughes2018inequity,yang2020learning,jiang2019learning}, agents are not able to punish each other in the traffic light control problem and self-punishment of an agent does not encourage the agent to punish others. This may partly explain the difference between the claimed near-optimal hyperparameters. 

The results highlight the potential of reshaped rewards in improving the performance of deep reinforcement learning methods. Moreover, the higher performance of the IA model for the newly tested range of parameters suggests performing a search over the coefficients $\alpha$ and $\beta$ of the inequities in future applications. Automatic search on these parameters is subject to future studies. 

\backmatter

\bmhead{Acknowledgments}
We would like to thank Digital Research Alliance of Canada for providing computational resources that facilitated our experiments.
 
\section*{Declarations}

\begin{itemize}
\item Funding \\
No funds, grants, or other support was received.
\item Competing interests \\
The authors declare no competing interest.
\item Ethics approval and Consent to participate/publication \\
`` Not applicable''
\item Code/data availability  \\
The codes are available online at \href{https://github.com/MersadHJ/IACoLight}{ https://github.com/MersadHJ/IACoLight}.
\item Authors' contributions \\
MH contributed to the algorithm and experiments. FA 
contributed to the idea and analysis, and took the lead on the writing. PR contributed to the idea and analysis. FA and PR supervised the study. All authors contributed to the writing.
\end{itemize}

\bibliography{mybib}

\end{document}